# Cryogenic temperatures as a path towards high-*Q* terahertz metamaterials

**Ranjan Singh,[1] Zhen Tian,[1,2] Jiaguang Han,[3] Carsten Rockstuhl,[4] Jianqiang Gu,[1,2] and Weili Zhang[1,a)]**

[1]*School of Electrical and Computer Engineering,*

*Oklahoma State University, Stillwater, Oklahoma 74078, USA*

[2]*Center for Terahertz waves and College of Precision Instrument and Optoelectronics Engineering, Tianjin University, and the Key Laboratory of Optoelectronics Information and Technical Science (Ministry of Education), Tianjin 300072, People's Republic of China*

[3]*Department of Physics, National University of Singapore, 2 Science Drive 3, Singapore*

[4] *Institute of Condensed Matter Theory and Solid State Optics, Friedrich-Schiller-Universität Jena, Jena 07743, Germany*

## Abstract

Optical properties of planar thin film metamaterials were measured at room and liquid nitrogen temperatures using terahertz time-domain spectroscopy. The operation of metamaterials at cryogenic temperatures is anticipated to be a promising path towards low-loss metamaterials since nonradiative losses are strongly suppressed due to higher charge mobility. A 14% increase in the quality factor of the resonances was experimentally observed. It was limited by the high electron scattering rate due to defects in thin films. Supplementary simulations assuming metamaterials made of thick films reveal a temperature controlled behaviour and a 40% increase of the *Q*-factor at 10 K.

[a)]Electronic mail: weili.zhang@okstate.edu

The exciting field of metamaterials (MMs) has seen an unprecedented growth over the past decade. Canonical unit elements of MMs can be scaled in size to operate across many decades of the electromagnetic spectrum. They all share the ability to engineer the propagation characteristics of electromagnetic waves in a way that cannot be achieved with naturally occurring materials. Thus, MMs have found a variety of applications. Examples are, but are not limited to, optical magnetism **[1-3]**, sensing devices **[4,5]**, asymmetric transmission **[6,7]**, and frequency selective surfaces. Planar MMs, or metafilms, are made of thin metal films with a thickness comparable to the skin depth. Basically they constitute subwavelength patterns on dielectric substrates. The resonances sustained by these patterns are at the heart of any MM and any application mentioned above. In spite of several newly found potential applications of these metafilms, practical use is often hindered by strong losses of the metallic elements. It causes a weakening and a damping of the resonance. This makes dissipation the property that dominates the light propagation and not the dispersive effects which are usually at the focus of interest. Therefore, the compensation of such losses is currently the most urgent topic one needs to solve prior lifting the topic of MMs to a new stage **[8]**. The losses itself have a radiative contribution, which can be tailored by suitable geometrical modifications **[9]**, and a non-radiative contribution, associated to the intrinsic absorption of the light in the metal.  Consequently, in order to improve their performance, the conductivity of the metafilms must be increased. A simple technique to increase the metal conductivity is to cool it to liquid nitrogen or helium temperatures **[10]**.

Here we investigate the behavior of terahertz planar MM at cryogenic temperatures. It is shown that thin film MMs show a 14% increase of the quality (*Q*-) factor of the inductive-capacitive (LC) resonance at liquid nitrogen temperature. Supplementary simulations about the optical properties of planar MMs made of thick metal films reveal an almost 40% increase of the *Q*-factor at the same resonance. The transmission level can be engineered over a dynamic range of 15dB. Distinction between thin and thick metafilms constitutes an opportunity to investigate the fundamental mechanisms that affect the temperature dependent damping of the resonances in terahertz MMs. In perspective, the chosen approach provides an alternative route to reduce the Ohmic losses in the metafilm split ring resonators (SRRs). This may lead to the development of temperature controlled terahertz devices and components.

The cooling experimental setup is shown in Fig. 1(a). The MM sample and the reference sample are screwed evenly on a metallic sample holder and the assembly is placed in a vacuum chamber with straight through optical access well positioned at the center of the standard THz-TDS system **[11,12]**. The vacuum chamber has a liquid nitrogen container in tight contact with the MM sample so that effective cooling of the MM is possible. In the 8*f* confocal THz-TDS set-up the transmitter is a GaAs photoconductive switch gated by the femto-second laser beam with a repetition rate of 88 MHz and the detector is a ZnTe module. The terahertz beam size at focus is 2 mm and the S/N ratio is about 3000:1. The MM is formed by planar array of pairs of concentric subwavelength SRRs as shown in Fig. 1(b) **[13]**. The double SRR metafilm pattern was fabricated using photolithographic technique and then depositing 150 nm of Aluminum metal on 640 μm p-type silicon substrate ($\varepsilon = 11.68$). The metafilm thickness is equivalent to 1.3 times the skin depth of Al at 0.5 THz. The unit cell of the square double SRR pattern has a periodicity of $P = 50$ μm. Other structural dimensions are $t = 6$ μm, $d = 2$ μm, and $l = 36$ μm, as shown in Fig. 1(c). The overall size of the MM sample is 15 mm × 15 mm and the 2 mm diameter terahertz beam excites several thousand SRRs at normal incidence with electric field polarization oriented along the SRR gaps. An identical blank silicon wafer with no patterns is used as a reference.

Figure 2 shows the measured transmission response of the thin planar MM sample illuminated at normal incidence at room and at liquid nitrogen temperature. The delay in Fig. 2(a) between the time domain sub-picoseond pulses is due to temperature dependent refractive index of Si ($dn/dT = 1.3 \times 10^{-4}$/K) **[10]**. The transmission spectrum in Fig. 2(b) is obtained by normalizing the measured transmission to the reference transmission of a blank p-type silicon wafer identical to the sample substrate. It is defined as $\left|E_{\mathrm{s}}(\omega)/E_{\mathrm{r}}(\omega)\right|$, where $E_{\mathrm{s}}(\omega)$ and $E_{\mathrm{r}}(\omega)$ are Fourier transformed time traces of the transmitted electric fields of the signal and the reference pulses, respectively. With the incident terahertz field polarization being along the SRR gaps, the odd eigenmode of the MM are excited **[14]**. The lowest order odd eigenmode is usually coined as Inductive-Capacitive (LC) resonance. The LC resonance is of particular importance to affect the dispersion in the effective permittivity at normal incidence and the effective permeability at an in-plane illumination. Therefore, we will focus mainly on the behavior of this resonance mode with respect to the reduced temperature. The higher frequency resonance at 1.6 THz is the second order odd eigenmode. The measured spectrum reveals line narrowing at liquid nitrogen temperature in both resonances. The LC resonance at 0.5 THz narrowed down by 19 GHz resulting in a 14% increase in the *Q*-factor. The transmission level drops by 5% at the LC resonance and gets enhanced over wide ‘off resonance’ bandwidth between 0.5 to 1.6 THz.

The origin of LC resonance can be attributed to the excitation of circular current by the incident electric field inducing magnetic moment perpendicular to the MM plane. The strength of the current in the rings determines the sharpness of the resonance **[15-19]**. Each double SRR can be modeled as a series RLC circuit. The Ohmic resistance of the metallic SRRs is directly related to the circulating current, and plays a key role in describing the damping of these subwavelength resonators. The decrease in effective Ohmic resistance $R_l$ of the SRRs is mainly responsible for the LC resonance sharpening when the MM is subjected to cryogenic temperatures since the metal conductivity can be increased at reduced temperatures **[10]**. The effective resistance of the SRRs can be estimated by the equivalent ring model for the current distribution **[17-19]**. For the square double SRRs, the effective resistance is approximately given as $R_l = 4l' / (th\sigma_{dc})$ when $h < 2\delta$ holds, and $R_l = 2l' / (t\delta\sigma_{dc})$ when $h \geq 2\delta$ holds, $\delta$ being the skin depth of the metafilm, $l'$ = 21 µm is the average side length of the SRRs, *t* is the width of the metal lines, and *h* and $\sigma_{dc}$ being the thickness and d.c. conductivity of the SRR film, respectively. The dc conductivity is given by $\sigma_{dc} = \varepsilon_0 \omega_p^2 / \gamma$, where $\gamma$ is the collision frequency or the scattering rate of the electrons. The Matthiessen's rule defines the total scattering $\gamma = \gamma_P + \gamma_D$ as the sum of temperature dependent term $\gamma_P$, dominated by scattering losses of electrons to phonons and a temperature independent term $\gamma_D$, dominated by scattering losses of electrons at defects in the lattice of the metal film. The imperfections in the lattice crystal are due to impurities and various other defects such as vacancies, grain boundaries, stacking faults and dislocations **[20]**. In thin metal films with a thickness comparable to the skin depth the losses via electron scattering at defects dominate. Therefore the increase in the conductivity is approximately a factor of two **[10]**. By contrast, a thick metal film undergoes an order of magnitude increase in the conductivity values when cooled to liquid nitrogen temperature **[21]**.

Since this regime is currently experimentally not accessible to us, we simulated the low temperature effect in the MMs made of thick metallic films to reveal how the optical properties would be affected if metal films of a strongly reduced defect density would be used. In the simulations we kept every dimension of the double SRR identical except the film thickness, *h* was increased to 10 µm. We wish to note that the fabrication of MMs with such a thickness is possible by using proton beam writing process **[22]**. Simulations were made with CST Microwave Studio**.** In the simulations we assumed a temperature dependent conductivity for the Al metal as documented in the literature **[21]**. Figure 4(a) shows for illustrative purpose the tremendous change in bulk Al metal conductivity at temperatures lower than 77 K. Figure 3 shows the transmission spectrum of the thick film MM as the temperature is reduced to very low values. The MM shows high modulation with temperature as the transmission level gradually reduced from -17 dB at room temperature to -32 dB at 10 K. The transmission reveals saturation at 10 K. Figure 4a shows the decrease in the linear amplitude transmission depending on the temperature at the resonance frequency. Figure 4(b) shows the gradual strengthening of LC resonance in terms of $Q$-factor from $Q$ = 7.6 to 10.5 with decreasing temperature. It can be clearly seen that as the Ohmic effective resistance decreases, there is an increasing current in the RLC circuit which results in 40% higher $Q$-factor. It is important to note that even though the Ohmic resistance reaches zero at 10 K, the $Q$-factor does not increase indefinitely since the subwavelength double SRR has a sizable radiation resistance which is not affected by the present approach. Only a further adjustment of the geometry of the unit cell would allow affecting these radiative losses.

In conclusion, we demonstrate the low temperature behavior of planar terahertz MMs. An increase of the quality factor of thin film planar MM by 14% was experimentally show to be possible since the conductivity of the metal films increases; however, it is limited by the scattering of electrons at defects in the skin depth scale lattice crystal. Such defects are grain boundaries, stacking faults and dislocations. Based on simulations it is anticipated, that the problem can be lifted by using thick metal films. Since the density of defects in such bulk materials is strongly reduced, the conductivity at 77 K will be an order higher when compared to room temperature. Such thick film planar MMs show an overall resonance sharpening of 40% from room temperature to 10 K in a switchable fashion. Below 10 K, the optical properties tend to be temperature independent and the remaining line shape is predominantly dictated by radiative damping and a temperature independent component of the electron scattering rate. MMs at cryogenic temperatures have the potential to be used as low loss structures across all the domains of the electromagnetic spectrum. The tunability achieved here would open up avenues for the development of temperatures controlled terahertz devices and components.

This work was partially supported by the US National Science Foundation, the China scholarship council, and the German Federal Ministry of Education and Research (Metamat).

**References**

1. J. B. Pendry, A. Holden, D. Robbins, W. Stewart, IEEE Trans. Microwave Theory Tech **7**, 2075, (1999).
2. R. A. Shelby, D. R. Smith, and S. Schultz, Science **292,** 5514 (2001).
3. T. J. Yen, W. J. Padilla, N. Fang, D. C. Vier, D. R. Smith, J. B. Pendry, D. N. Basov, and X. Zhang, Science **303**, 1494, (2004).
4. J. F. O'Hara, R. Singh, I. Brener, E. Smirnova, J. Han, A. J. Taylor, and W. Zhang, Opt. Express **16**, 1786 (2008).
5. I. A. I. A-Naib, C. Jansen, and M. Koch, App. Phys. Lett. **93**, 083507 (2008).
6. V. A. Fedotov, P. L. Mladyonov, S. L. Prosvirnin, A. V. Rogacheva, Y. Chen, and N. I. Zheludev, Phys. Rev. Lett. **97**, 167401(2006).
7. R. Singh, E. Plum, C. Menzel, C. Rockstuhl, A. K. Azad, R. A. Cheville, F. Lederer, W. Zhang, and N. I. Zheludev, Phys. Rev B **80**, 153104 (2009).
8. S. A. Ramakrishna and J. B. Pendry, Phys. Rev. B **67**, 201101 (2003).
9. T. Hao, C.J. Stevens, and D.J. Edwards, Electron. Lett. **41**, 653 (2005).
10. N. Laman, and D. Grischkowsky, App. Phys. Lett. **93**, 051105 (2008).
11. D. Grischkowsky, S. Keiding, M. van Exter, and **C**h. Fattinger, J. Opt. Soc. Am. B **7**, 2006 (1990).
12. Q. W. M. Litz, and X. C. Zhang, App. Phys. Lett. **68**, 21 (1996).
13. A. K. Azad, J. Dai, and W. Zhang, Opt. Lett. **31**, 634 (2006).
14. C. Rockstuhl, T. Zentgraf, T. P. Meyrath, H. Giessen, and F. Lederer, Opt. Express **16**, 2080 (2008).

15. W. J. Padilla, A. J. Taylor, C. Highstrete, M. Lee, R. D. Averitt, Phys. Rev. Lett. **96**, 107401 (2006).

16. H. T. Chen, W. J. Padilla, J. M. O. Zide, A. C. Gossard, A. J. Taylor, and R. D. Averitt, Nature **444**, 597 (2006).

17. R. Marques,F. Mesa, Jesus Martel and F. Median, IEEE Transactions on Antennas and Propagation **51**, 2572 (2003).

18. R. Singh, E. Smirnova, A. J. Taylor, J. F. O'Hara, and W. Zhang, Opt. Express **16**, 6537 (2008).

19. R. Singh, A. K. Azad, J. F. O'Hara, A. J. Taylor, and W. Zhang, Opt. Lett. **33**, 1506 (2008).

20. A. B. Pippard, *The Dynamics of Conduction Electrons* (Gordan and Breach Science, New York, 1965)

21. *CRC handbook of Chemistry and Physics*, 87th ed., edited by D. R. Lide (CRC, Boca Raton, FL, 2006), pp. 12-39.

22. S. Y. Chiam, R. Singh, J. Gu, J. Han, W. Zhang, and A. A. Bettiol, Appl. Phys. Lett. **94**, 064102 (2009).

**Figure Captions**

FIG. 1 (color online). (a) The THz-TDS setup with a standard cryostat cooling the sample, S and the reference, R to liquid nitrogen temperatures. (b) Microscopic image of the sample array. (c) Unit cell with dimension parameters, $t$ = 6 μm, $d$ = 2 μm, $l$ = 36 μm, $l'$ = 21 μm, $h$ = 150 nm. The periodicity of the unit cells is $P$ = 50 μm

FIG. 2 (color online). Measured (a) sub-picosecond transmitted pulses through the metamaterial and (b) amplitude transmission at room and liquid nitrogen temperatures. The incident E field is polarized parallel to the gap of the double SRRs.

FIG. 3 (color online). Simulated amplitude transmission spectra of the bulk film ($h$ = 10 μm) MM at decreasing temperatures.

FIG. 4 (color online). (a) Handbook conductivity values of bulk Al metal at reducing temperatures and the change in amplitude transmission with increasing conductivity of the bulk film MM, (b) extracted $Q$-factor and the calculated effective Ohmic resistance of bulk film MM with decrease in the temperature; Blue and red curves are just to guide the eye.

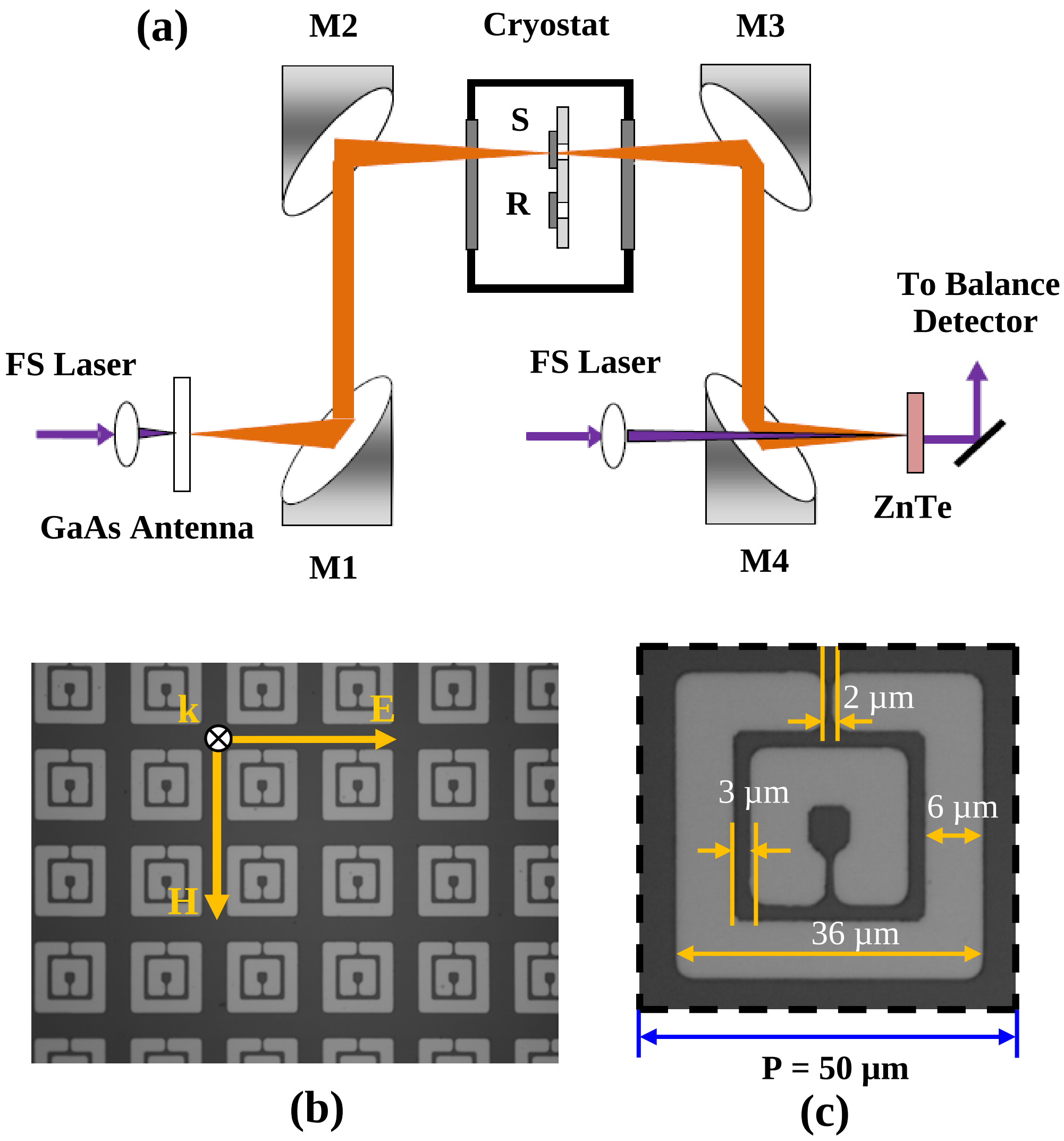

**Figure 1.**
**Singh *et al.***

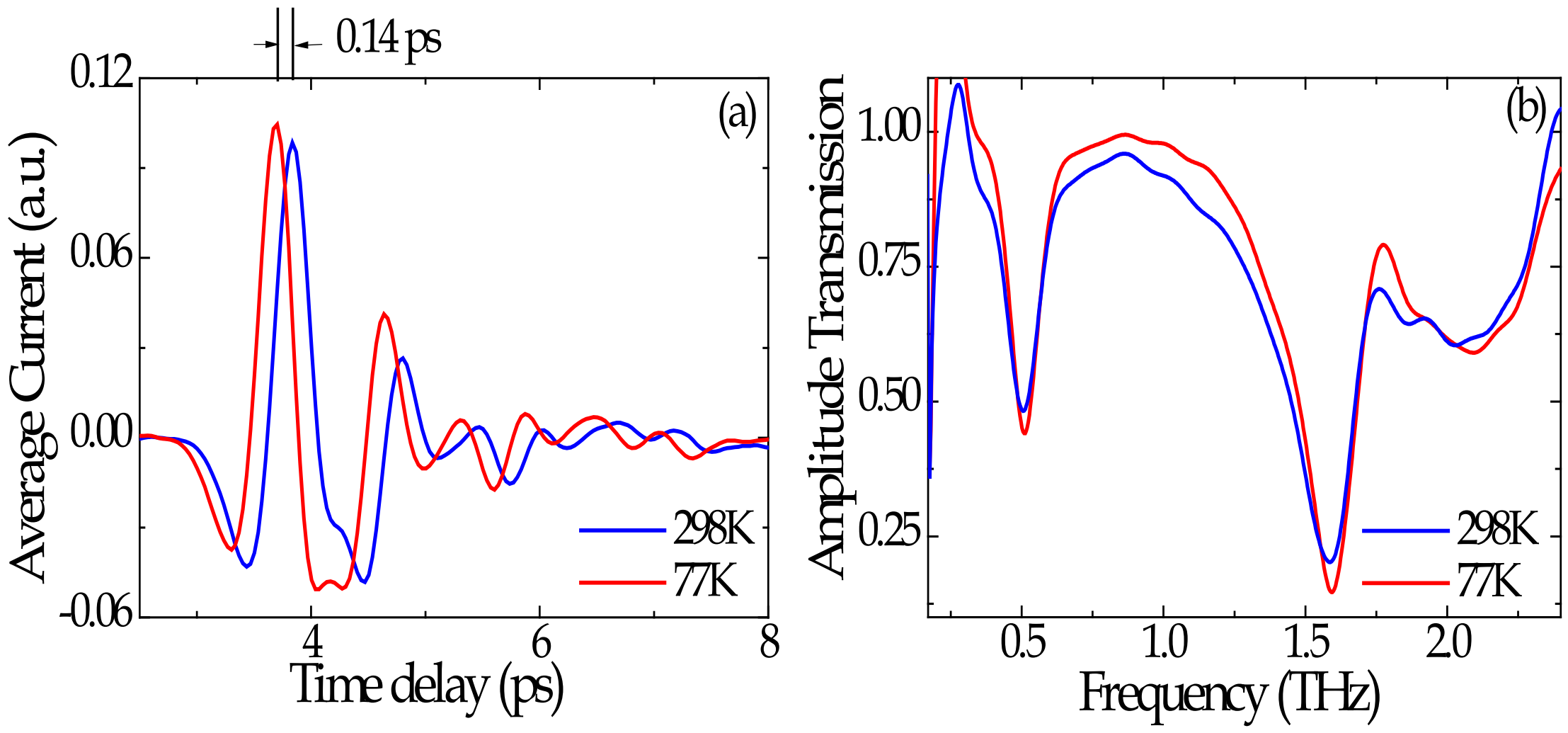

**Figure 2.**
**Singh *et al.***

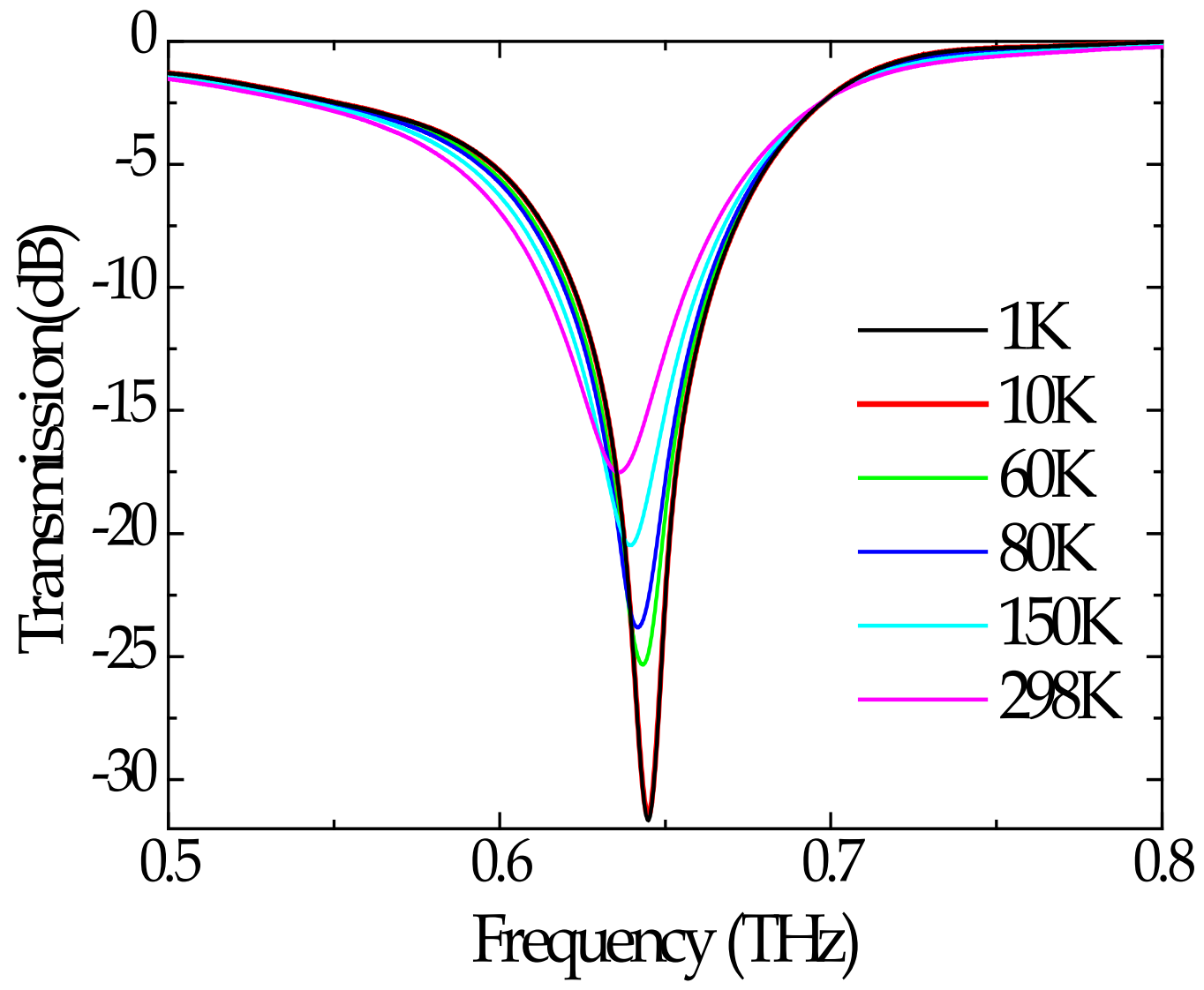

**Figure 3.**
**Singh *et al.***

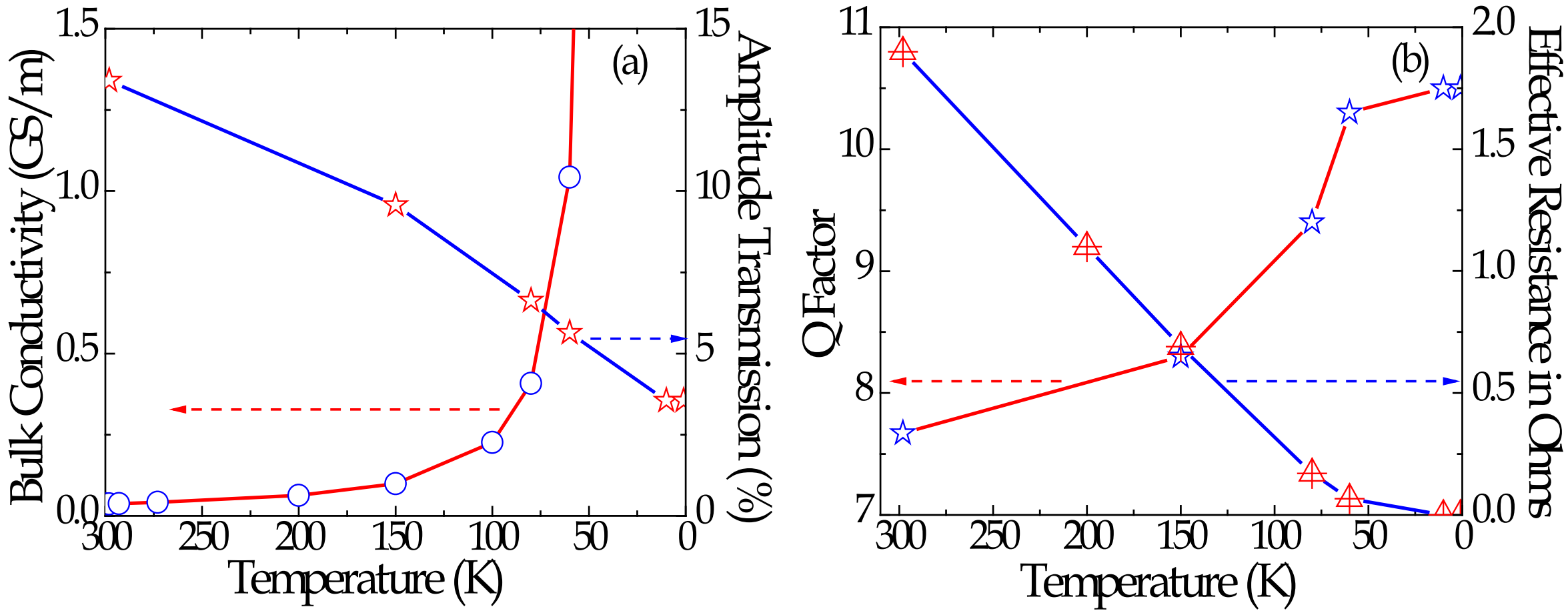

**Figure 4.**
**Singh *et al.***